\documentstyle[12pt]{article}

\def\title#1{\begin{center}{\Large\bf #1}\end{center}}
\def\author#1{\vskip 5mm \begin{center}{#1}\end{center}}
\def\address#1{\begin{center}{\it #1}\end{center}}

\newcommand{\beq}{\begin{equation}}
\newcommand{\eeq}{\end{equation}}
\newcommand{\bea}{\begin{eqnarray}}
\newcommand{\eea}{\end{eqnarray}}

\begin{document}

\title{Optimal supply against fluctuating demand}
\author{Nobuyuki Sakai
\footnote{Electronic address: sakai@ke-sci.kj.yamagata-u.ac.jp}
and Hisanori Kudoh}
\address{Faculty of Education, Yamagata University, Yamagata 990-8560, Japan}
\date{23 March 2005}

\begin{abstract}
Sornette et al.\ claimed that the optimal supply does not agree with the
average demand, by analyzing a bakery model where a daily demand
fluctuates with a uniform distribution.
In this note, we extend the model to general probability
distributions, and obtain the formula of the optimal supply for
Gaussian distribution, which is more realistic. Our result is useful
in a real market to earn the largest income on average.

\vskip 5mm\noindent
keyword: demand, supply, fluctuations
\end{abstract}

\baselineskip = 24pt

\section{Introduction}

Sornette et al.\ (1999) claimed that the optimal supply does
not agree with the average demand, contrary to the common sense in
economy. They considered a bakery model where a daily demand
fluctuates with a uniform distribution, and derived the formula of the
optimal supply. Although their result is reasonable and
meaningful, it is not clear how the result will change if we consider
a different distribution.
In this note, we extend the model to general probability
distributions, and calculate the optimal supply for the
Gaussian distribution, which is more realistic in a market.

\section{Model and analysis of Sornette, Stauffer and Takayasu}

In this section we review the model and analysis of Sornette
et al.\  Let us consider a bakery shop where baked croissants are sold
every day. A question is how many croissants should be baked a day to
make the maximal profit.
We define the variables as follows.
\begin{itemize}
\item $x$: the selling price of a croissant.
\item $y$: its production cost.
\item $s$: the production number of croissants per day (supply).
\item $n$: the number of croissants requested by customers per day
(demand).
\item $D$: the average demand, i.e., $D\equiv<n>$.
\end{itemize}
The expectation of the total profit $L(s)$ is given by
\beq\label{L}
L(s)\equiv<x~{\rm min}(n,s)-ys> 
=x\int^s_0nP(n)dn+xs\int^{\infty}_sP(n)dn-ys,
\eeq
where $P(n)$ is the probability distribution of $n$.

Sornette et al.\ assumed, for simplicity, a uniform distribution,
\beq\label{uniform}
P_u(n)\equiv
\left\{\begin{array}{ll}
1/2\delta ~& {\rm for} ~~~D-\delta\le n\le D+\delta\\
0         ~& {\rm for} ~~~n<D-\delta,~D+\delta<n.
\end{array}\right.
\eeq
Then one can integrate (\ref{L}) as
\beq
L(s)=-{x\over4\delta}\left\{s-D-\delta\left(1-{2y\over
x}\right)\right\}^2
+(x-y)\left(D-{\delta y\over x}\right).
\eeq
$L(s)$ takes the maximam value when $s$ takes
\beq\label{su}
s_{{\rm max}}\equiv D+\delta\left(1-{2y\over x}\right).
\eeq
This shows that, if the cost per price, $y/x$, is larger (smaller)
than half, the optimal demand, $s_{{\rm max}}$, is smaller (larger)
than the average demand, $D$.

\section{Optimal supply for Gaussian distribution}

Let us re-analyze (\ref{L}) for general probability distributions. We
do not have to integrate (\ref{L}) directly, because what we want to
know is the optimal supply $s_{{\rm max}}$, which is given by
\beq\label{dL}
{dL\over ds}(s_{{\rm max}})=x\int^{\infty}_{s_{{\rm max}}}P(n)dn-y=0.
\eeq
This simple equation gives the optimal supply for general probability
distributions.

If we assume Gaussian distribution,
\beq
P_G(n)\equiv{1\over\sqrt{2\pi}\sigma}
\exp\left[-{(n-D)^2\over2\sigma^2}\right],
\eeq
the above integration is expressed as
\beq
\int^{\infty}_{s_{{\rm max}}}P_G(n)dn=\frac12-\frac12
{\rm Erf}\left({s_{{\rm max}}-D\over\sqrt{2}\sigma}\right),
\eeq
where Erf is the error function, which is defined as
\beq
{\rm Erf}~z\equiv{2\over\sqrt{\pi}}\int^z_0e^{-t^2}dt.
\eeq
Then we arrive at the formula of the optimal supply for the Gaussian
distribution,
\bea\label{sG}
{s_{{\rm max}}-D\over\sigma}
&=&\sqrt{2}{\rm Erf}^{-1}\left(1-\frac{2y}x\right) \nonumber\\
&=&\sqrt{{\pi\over2}}\left(1-\frac{2y}x\right)
+{\sqrt{2}\pi^{\frac32}\over24}\left(1-\frac{2y}x\right)^3
+O\left[\left(1-\frac{2y}x\right)^5\right]
~~~({\rm Gaussian}).
\eea
Because the cost is usually in the range $0.3x<y<0.7x$, which reads
$|1-2y/x|<0.4$, the first-order approximation in
(\ref{sG}) is sufficient in most cases.

For reference, we rewrite the result for the uniform distribution
(\ref{su}). Because the variance of the uniform distribution (\ref{uniform}) is 
evaluated as $\sigma^2=\delta^2/3$, (\ref{su}) is rewritten as
\beq\label{su2}
{s_{{\rm max}}-D\over\sigma}=\sqrt{3}\left(1-\frac{2y}x\right)
~~~({\rm uniform}).
\eeq

We see that the difference between $\sqrt{\pi/2}\approx1.25$ in
(\ref{sG}) and $\sqrt{3}\approx1.73$ in (\ref{su2}) is not negligible.
Contrary to the speculation of
Sornette {\it et al.}, however, the critical value of the
cost-to-price ratio, $y/x=1/2$, is unchanged.
Because Gaussian distribution is more realistic, 
our simple formula (\ref{sG}) is useful
in a real market to earn the largest income on average.


\end{document}